# Foundations of Electromagnetism, Equivalence Principles and Cosmic Interactions

Wei-Tou Ni

*Center for Gravitation and Cosmology*
*Department of Physics, National Tsing Hua University*
*Hsinchu, Taiwan, 300 ROC,*
*and*
*Shanghai United Center for Astrophysics*
*Shanghai Normal University*
*Shanghai, 200234 China*

## 1. Introduction

Standard electromagnetism is based on Maxwell equations and Lorentz force law. It can be derived by a least action with the following Lagrangian density for a system of charged particles in Gaussian units (e.g., Jackson, 1999),

$$L_{EMS} = L_{EM} + L_{EM\text{-}P} + L_P = -(1/(16\pi))[(1/2)\eta^{ik}\eta^{jl} - (1/2)\eta^{il}\eta^{kj}]F_{ij}F_{kl} - A_k j^k - \Sigma_I m_I[(ds_I)/(dt)]\delta(\mathbf{x}-\mathbf{x}_I), \qquad (1)$$

where $F_{ij} \equiv A_{j,i} - A_{i,j}$ is the electromagnetic field strength tensor with $A_i$ the electromagnetic 4-potential and comma denoting partial derivation, $\eta^{ij}$ is the Minkowskii metric with signatute (+, -, -, -), $m_I$ the mass of the $I$th charged particle, $s_I$ its 4-line element, and $j^k$ the charge 4-current density. Here, we use Einstein summation convention, i.e., summation over repeated indice. There are three terms in the Lagrangian density $L_{EMS}$ – (i) $L_{EM}$ for the electromagnetic field, (ii) $L_{EM\text{-}P}$ for the interaction of electromagnetic field and charged particles and (iii) $L_P$ for charged particles.

The electromagnetic field Lagrangian density can be written in terms of the electric field **E** [≡ $(E_1, E_2, E_3) \equiv (F_{01}, F_{02}, F_{03})$] and magnetic induction **B** [≡ $(B_1, B_2, B_3) \equiv (F_{32}, F_{13}, F_{21})$] as

$$L_{EM} = (1/8\pi)[\mathbf{E}^2 - \mathbf{B}^2]. \qquad (2)$$

This classical Lagrangian density is based on the photon having zero mass. To include the effects of nonvanishing photon mass $m_{photon}$, a mass term $L_{Proca}$,

$$L_{Proca} = (m_{photon}^2 c^2 / 8\pi\hbar^2)(A_k A^k), \qquad (3)$$

needs to be added (Proca, 1936a, 1936b, 1936c, 1937, 1938). We use $\eta^{ij}$ and its inverse $\eta_{ij}$ to raise and lower indices. With this term, the Coulomb law is modified to have the electric potential $A_0$,

$$A_0 = q(e^{-\mu r}/r), \qquad (4)$$

where $q$ is the charge of the source particle, $r$ is the distance to the source particle, and $\mu$ ($\equiv m_{photon} c/\hbar$) gives the inverse range of the interaction.

Experimental test of Coulomb's law (Williams, Faller & Hill, 1971) gives a constraint of photon mass as

$$m_{photon} \leq 10^{-14} \text{ eV } (= 2 \times 10^{-47} \text{ g}), \qquad (5)$$

or the interaction range $\mu^{-1}$ as

$$\mu^{-1} \geq 2 \times 10^7 \text{ m.} \qquad (6)$$

Photon mass affects the structure and the attenuation of magnetic field and therefore can be constrained by measuring the magnetic field of Earth, Sun or an astronomical body (Schrödinger, 1943; Bass & Schrödinger, 1955). From the magnetic field measurement of Jupiter during Pioneer 10 flyby, constraints are set as (Davis, Goldhaber & Nieto, 1975)

$$m_{photon} \leq 4 \times 10^{-16} \text{ eV } (= 7 \times 10^{-49} \text{ g}); \; \mu^{-1} \geq 5 \times 10^8 \text{ m.} \qquad (7)$$

Using the plasma and magnetic field data of the solar wind, constraints on photon mass are set recently as (Ryutov, 2007)

$$m_{photon} \leq 10^{-18} \text{ eV } (= 2 \times 10^{-51} \text{ g}); \; \mu^{-1} \geq 2 \times 10^{11} \text{ m.} \qquad (8)$$

Large-scale magnetic fields in vacuum would be direct evidence for a limit on their exponential attenuation with distance, and hence a limit on photon mass. Using observations on galactic sized fields, Chibisov limit is obtained (Chibisov, 1976)

$$m_{photon} \leq 2 \times 10^{-27} \text{ eV } (= 4 \times 10^{-60} \text{ g}); \; \mu^{-1} \geq 10^{20} \text{ m.} \qquad (9)$$

For a more detailed discussion of this work and for a comprehensive review on the photon mass, please see Goldhaber and Nieto (2010).

As larger scale magnetic field discovered and measured, the constraints on photon mass and on the interaction range may become more stringent. If cosmic scale magnetic field is discovered, the constraint on the interaction range may become bigger or comparable to Hubble distance (of the order of radius of curvature of our observable universe). If this happens, the concept of photon mass may lose significance amid gravity coupling or curvature coupling of photons.

Now we turn to quantum corrections to classical electrodynamics. In classical electrodynamics, the Maxwell equations are linear in the electric field **E** and magnetic field **B**, and we have the principle of superposition of electromagnetic field in vacuum. However, in the electrodynamics of continuous matter, media are usually nonlinear and the principle of superposition of electromagnetic field is not valid. In quantum electrodynamics, due to loop diagrams like the one in Fig. 1, photon can scatter off photon in vacuum. This is the origin of invalidity of the principle of superposition and makes vacuum a nonlinear medium also. The leading order of this effect in slowly varying electric and magnetic field is derived in Heisenberg and Euler (1936) and can be incorporated in the Heisenberg-Euler Lagrangian density

$$L_{Heisenberg-Euler} = [2\alpha^2\hbar^2/45(4\pi)^2 m^4 c^6][(\mathbf{E}^2-\mathbf{B}^2)^2 + 7(\mathbf{E}\cdot\mathbf{B})^2], \qquad (10)$$

where α is the fine structure constant and *m* the electron mass. In terms of critical field strength $B_c$ defined as

$$B_c \equiv E_c \equiv m^2c^3/e\hbar = 4.4\times10^{13}\ G = 4.4\times10^9\ T = 4.4\times10^{13}\ statvolt/cm = 1.3\times10^{18}\ V/m, \qquad (11)$$

this Lagrangian density can be written as

$$L_{Heisenberg-Euler} = (1/8\pi)\ B_c^{-2}\ [\eta_1(\mathbf{E}^2-\mathbf{B}^2)^2 + 4\eta_2(\mathbf{E}\cdot\mathbf{B})^2], \qquad (12)$$

$$\eta_1 = \alpha/(45\pi) = 5.1\times10^{-5}\ \text{and}\ \eta_2 = 7\alpha/(180\pi) = 9.0\times10^{-5}. \qquad (13)$$

For time varying and space varying external fields, and higher order corrections in quantum electrodynamics, please see Dittrich and Reuter (1985) and Kim (2011) and references therein.

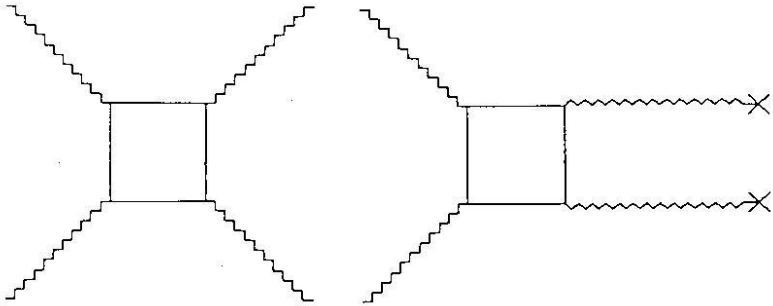

Fig. 1. On the left is the basic diagram for light-light scattering and for nonlinear electrodynamics; on the right is the basic diagram for the nonlinear light (electromagnetic-wave) propagation in strong electric and/or magnetic field.

Before Heisenberg & Euler (1936), Born and Infeld (Born, 1934; Born & Infeld, 1934) proposed the following Lagrangian density for the electromagnetic field

$$L_{Born\text{-}Infeld} = -(b^2/4\pi) \, [1 - (\mathbf{E}^2-\mathbf{B}^2)/b^2 - (\mathbf{E}\cdot\mathbf{B})^2/b^4]^{1/2}, \tag{14}$$

where $b$ is a constant which gives the maximum electric field strength. For field strength small compared with $b$, (14) can be expanded into

$$L_{Born\text{-}Infeld} = (1/8\pi) \, [(\mathbf{E}^2-\mathbf{B}^2) + (\mathbf{E}^2-\mathbf{B}^2)^2/b^2 + (\mathbf{E}\cdot\mathbf{B})^2/b^2 + O(b^{-4})]. \tag{15}$$

The lowest order of Born-Infeld electrodynamics agrees with the classical electrodynamics. The next order corrections are of the form of Eq. (12) with

$$\eta_1 = \eta_2 = B_c^2/b^2. \tag{16}$$

In the Born-Infeld electrodynamics, $b$ is the maximum electric field. Electric fields at the edge of heavy nuclei are of the order of $10^{21}$ V/m. If we take $b$ to be $10^{21}$ V/m, then, $\eta_1 = \eta_2 = 5.9 \times 10^{-6}$.

For formulating a phenomenological framework for testing corrections to Maxwell-Lorentz classical electrodynamics, we notice that $(\mathbf{E}^2-\mathbf{B}^2)$ and $(\mathbf{E}\cdot\mathbf{B})$ are the only Lorentz invariants second order in the field strength, and $(\mathbf{E}^2-\mathbf{B}^2)^2$, $(\mathbf{E}\cdot\mathbf{B})^2$ and $(\mathbf{E}^2-\mathbf{B}^2)(\mathbf{E}\cdot\mathbf{B})$ are the only Lorentz invariants fourth order in the field strength. However, $(\mathbf{E}\cdot\mathbf{B})$ is a total divergence and, by itself in the Lagrangian density, does not contribute to the equation of motion (field equation). Multiplying $(\mathbf{E}\cdot\mathbf{B})$ by a pseudoscalar field $\Phi$, the term $\Phi(\mathbf{E}\cdot\mathbf{B})$ is the Lagrangian density for the pseudoscalar-photon (axion-photon) interaction. When this term is included together with the fourth-order invariants, we have the following phenomenological Lagrangian density for our Parametrized Post-Maxwell (PPM) Lagrangian density including various corrections and modifications to be tested by experiments and observations,

$$L_{PPM} = (1/8\pi)\{(\mathbf{E}^2-\mathbf{B}^2)+\xi\Phi(\mathbf{E}\cdot\mathbf{B})+B_c^{-2}[\eta_1(\mathbf{E}^2-\mathbf{B}^2)^2+4\eta_2(\mathbf{E}\cdot\mathbf{B})^2+2\eta_3(\mathbf{E}^2-\mathbf{B}^2)(\mathbf{E}\cdot\mathbf{B})]\}. \tag{17}$$

This PPM Lagrangian density contains 4 parameters $\xi$, $\eta_1$, $\eta_2$ & $\eta_3$, and is an extension of the two-parameteer ($\eta_1$ and $\eta_2$) post-Maxwellian Lagrangian density of Denisov, Krivchenkov and Kravtsov (2004). The manifestly Lorentz covariant form of Eq. (17) is

$$L_{PPM} = (1/(32\pi))\{-2F^{kl}F_{kl} -\xi\Phi F^{*kl}F_{kl}+B_c^{-2}[\eta_1(F^{kl}F_{kl})^2+\eta_2(F^{*kl}F_{kl})^2+\eta_3(F^{kl}F_{kl})(F^{*ij}F_{ij})]\}, \tag{18}$$

where

$$F^{*ij} \equiv (1/2)e^{ijkl} F_{kl}, \tag{19}$$

with $e^{ijkl}$ defined as

$$e^{ijkl} \equiv 1 \text{ if } (ijkl) \text{ is an even permutation of } (0123); -1 \text{ if odd}; 0 \text{ otherwise.} \tag{20}$$

In section 2, we derive the PPM nonlinear electrodynamic equations, and in section 3, we use them to derive the light propagation equation in PPM nonlinear electrodynamics. In section 4, we discuss ultra-high precision laser interferometry experiments to measure the parameters of PPM electrodynamics. In section 5, we treat electromagnetism in curved spacetime using Einstein Equivalence Principle, and discuss redshift as an application with examples from astrophysics and navigation. In section 6, we discuss empirical tests of electromagnetism in gravity and the χ-g framework and find pseudoscalar-photon interaction uniquely standing out. In section 7, we discuss the pseudoscalar-photon interaction and its relation to other approaches. In section 8, we use Cosmic Microwave Background (CMB) observations to constrain the cosmic polarization rotation and discuss radio galaxy observations. In section 9, we present a summary and an outlook briefly.

## 2. Equations for nonlinear electrodynamics

In analogue with the nonlinear electrodynamics of continuous media, we can define the electric displacement **D** and magnetic field **H** as follows:

$$\mathbf{D} \equiv 4\pi(\partial L_{PPM}/\partial \mathbf{E}) = [1 + 2\eta_1(\mathbf{E}^2 - \mathbf{B}^2)B_c^{-2} + 2\eta_3(\mathbf{E}\cdot\mathbf{B})B_c^{-2}]\mathbf{E} + [\Phi + 4\eta_2(\mathbf{E}\cdot\mathbf{B})B_c^{-2} + \eta_3(\mathbf{E}^2-\mathbf{B}^2)B_c^{-2}]\mathbf{B}, \quad (21)$$

$$\mathbf{H} \equiv -4\pi(\partial L_{PPM}/\partial \mathbf{B}) = [1 + 2\eta_1(\mathbf{E}^2 - \mathbf{B}^2)B_c^{-2} + 2\eta_3(\mathbf{E}\cdot\mathbf{B})B_c^{-2}]\mathbf{B} - [\Phi + 4\eta_2(\mathbf{E}\cdot\mathbf{B})B_c^{-2} + \eta_3(\mathbf{E}^2-\mathbf{B}^2)B_c^{-2}]\mathbf{E}. \quad (22)$$

From **D** & **H**, we can define a second-rank $G_{ij}$ tensor, just like from **E** & **B** to define $F_{ij}$ tensor. With these definitions and following the standard procedure in electrodynamics [see, e.g., Jackson (1999), p. 599], the nonlinear equations of the electromagnetic field are

$$\text{curl } \mathbf{H} = (1/c)\ \partial \mathbf{D}/\partial t + 4\pi\ \mathbf{J}, \quad (23)$$

$$\text{div } \mathbf{D} = 4\pi\ \rho, \quad (24)$$

$$\text{curl } \mathbf{E} = -(1/c)\ \partial \mathbf{B}/\partial t, \quad (25)$$

$$\text{div } \mathbf{B} = 0. \quad (26)$$

We notice that it has the same form as in macroscopic electrodynamics. The Lorentz force law remains the same as in classical electrodynamics:

$$d[(1 - \mathbf{v}_I^2/c^2)^{-1/2} m_I \mathbf{v}_I]/dt = q_I[\mathbf{E} + (1/c)\mathbf{v}_I \times \mathbf{B}] \quad (27)$$

for the $I$-th particle with charge $q_I$ and velocity $\mathbf{v}_I$ in the system. The source of $\Phi$ in this system is $(\mathbf{E}\cdot\mathbf{B})$ and the field equation for $\Phi$ is

$$\partial^i L_\Phi / \partial(\partial^i \Phi) - \partial L_\Phi / \partial \Phi = \mathbf{E}\cdot\mathbf{B}, \quad (28)$$

where $L_\Phi$ is the Lagrangian density of the pseudoscalar field $\Phi$.

## 3. Electromagnetic wave propagation in PPM electrodynamics

Here we follow the previous method (Ni et al., 1991; Ni, 1998), and separate the electric field and magnetic induction field into the wave part (small compared to external part) and external part as follows:

$$\mathbf{E} = \mathbf{E}^{wave} + \mathbf{E}^{ext}, \tag{29}$$

$$\mathbf{B} = \mathbf{B}^{wave} + \mathbf{B}^{ext}. \tag{30}$$

We use the following expressions to calculate the displacement field $\mathbf{D}^{wave}$ [= $(D^{wave}_a)$ = $(D^{wave}_1, D^{wave}_2, D^{wave}_3)$] and the magnetic field $\mathbf{H}^{wave}$ [= $(H^{wave}_a)$ = $(H^{wave}_1, H^{wave}_2, H^{wave}_3)$] of the electromagnetic waves:

$$D^{wave}_a = D_a - D^{ext}_a = (4\pi)[(\partial L_{PPM}/\partial E_a)_{E\&B} - (\partial L_{PPM}/\partial E_a)_{ext}], \tag{31}$$

$$H^{wave}_a = H_a - H^{ext}_a = -(4\pi)[(\partial L_{PPM}/\partial B_a)_{E\&B} - (\partial L_{PPM}/\partial B_a)_{ext}], \tag{32}$$

where $(\ldots)_{E\&B}$ means that the quantity inside paranthesis is evaluated at the total field values $\mathbf{E}$ & $\mathbf{B}$ and $(\ldots)_{ext}$ means that the quantity inside paranthesis is evaluated at the external field values $\mathbf{E}^{ext}$ & $\mathbf{B}^{ext}$.

Since both the total field and the external field satisfy Eqs. (23)-(26), the wave part also satisfy the same form of Eqs. (23)-(26) with the source terms subtracted:

$$\text{curl } \mathbf{H}^{wave} = (1/c) \ \partial \mathbf{D}^{wave}/\partial t, \tag{33}$$

$$\text{div } \mathbf{D}^{wave} = 0, \tag{34}$$

)

$$\text{curl } \mathbf{E}^{wave} = -(1/c) \ \partial \mathbf{B}^{wave}/\partial t, \tag{35}$$

$$\text{div } \mathbf{B}^{wave} = 0. \tag{36}$$

After calculating $D^{wave}_a$ and $H^{wave}_a$ from Eqs. (31) & (32), we express them in the following form:

$$D^{wave}_a = \Sigma_{\beta=1}^{3} \ \varepsilon_{\alpha\beta} \ E^{wave}_\beta + \Sigma_{\beta=1}^{3} \ \lambda_{\alpha\beta} \ B^{wave}_\beta, \tag{37}$$

$$H^{wave}_a = \Sigma_{\beta=1}^{3} \ (\mu^{-1})_{\alpha\beta} \ B^{wave}_\beta - \Sigma_{\beta=1}^{3} \ \lambda_{\beta\alpha} \ E^{wave}_\beta, \tag{38}$$

where

$$\varepsilon_{\alpha\beta}=\delta_{\alpha\beta}[1+2\eta_1(\mathbf{E}^2-\mathbf{B}^2)B_c^{-2}+2\eta_3(\mathbf{E}\cdot\mathbf{B})B_c^{-2}]+4\eta_1 E_\alpha E_\beta B_c^{-2}+4\eta_2 B_\alpha B_\beta B_c^{-2}+2\eta_3(E_\alpha B_\beta+E_\beta B_\alpha)B_c^{-2}, \tag{39}$$

$$(\mu^{-1})_{\alpha\beta} = \delta_{\alpha\beta}[1 + 2\eta_1(\mathbf{E}^2 - \mathbf{B}^2)B_c^{-2} + 2\eta_3(\mathbf{E}\cdot\mathbf{B})B_c^{-2}] - 4\eta_1 B_\alpha B_\beta B_c^{-2} - 4\eta_2 E_\alpha E_\beta B_c^{-2} + 2\eta_3(E_\alpha B_\beta + E_\beta B_\alpha)B_c^{-2}, \quad (40)$$

$$\lambda_{\alpha\beta} = \delta_{\alpha\beta}[\xi\Phi + 4\eta_2(\mathbf{E}\cdot\mathbf{B})B_c^{-2} + \eta_3(\mathbf{E}^2 - \mathbf{B}^2)B_c^{-2}] - 4\eta_1 E_\alpha B_\beta B_c^{-2} + 4\eta_2 B_\alpha E_\beta B_c^{-2} + 2\eta_3(E_\alpha E_\beta + B_\alpha B_\beta)B_c^{-2}, \quad (41)$$

and we have dropped the upper indices 'ext' for simplicity.

Using eikonal approximation, we look for plane-wave solutions. Choose the $z$-axis in the propagation direction. Solving the dispersion relation for $\omega$, we obtain

$$\omega_\pm = k\{1 + (1/4)[(J_1 + J_2) \pm [(J_1 - J_2)^2 + 4J^2]^{1/2}]\}, \quad (42)$$

where

$$J_1 \equiv (\mu^{-1})_{22} - \varepsilon_{11} - 2\lambda_{12}, \quad (43)$$

$$J_2 \equiv (\mu^{-1})_{11} - \varepsilon_{22} + 2\lambda_{21}, \quad (44)$$

$$J \equiv -\varepsilon_{12} - (\mu^{-1})_{12} + \lambda_{11} - \lambda_{22}. \quad (45)$$

Since the index of refraction $n$ is

$$n = k/\omega, \quad (46)$$

we find

$$n_\pm = 1 - (1/4)\{(J_1 + J_2) \pm [(J_1 - J_2)^2 + 4J^2]^{1/2}\}. \quad (47)$$

From this formula, we notice that "no birefringence" is equivalent to $J_1 = J_2$ and $J = 0$. A sufficient condition for this to happen is $\eta_1 = \eta_2$, $\eta_3 = 0$, and no constraint on $\xi$. We will show in the following that this is also a necessary condition. The Born-Infeld electrodynamics satisfies this condition and has no birefringence in the theory.

For $\mathbf{E}^{ext} = 0$, we now derive the refractive indices in the transverse external magnetic field $\mathbf{B}^{ext}$ for the linearly polarized lights whose polarizations (electric fields) are parallel and orthogonal to the magnetic field. First, we use Eqs. (39)-(41) & Eqs. (43)-(46) to obtain

$$\varepsilon_{\alpha\beta} = \delta_{\alpha\beta}[1 - 2\eta_1 \mathbf{B}^2 B_c^{-2}] + 4\eta_2 B_\alpha B_\beta B_c^{-2}, \quad (48)$$

$$(\mu^{-1})_{\alpha\beta} = \delta_{\alpha\beta}[1 - 2\eta_1 \mathbf{B}^2 B_c^{-2}] - 4\eta_1 B_\alpha B_\beta B_c^{-2}, \quad (49)$$

$$\lambda_{\alpha\beta} = \delta_{\alpha\beta}[\Phi - \eta_3 \mathbf{B}^2 B_c^{-2}] + 2\eta_3 B_\alpha B_\beta B_c^{-2}, \quad (50)$$

$$J_1 = -4\eta_1 B_2^2 B_c^{-2} - 4\eta_2 B_1^2 B_c^{-2} - 4\eta_3 B_1 B_2 B_c^{-2}, \quad (51)$$

$$J_2 = -4\eta_1 B_1^2 B_c^{-2} - 4\eta_2 B_2^2 B_c^{-2} + 4\eta_3 B_1 B_2 B_c^{-2}, \quad (52)$$

$$J = 4\eta_1 B_1 B_2 B_c^{-2} - 4\eta_2 B_1 B_2 B_c^{-2} + 2\eta_3(B_1^2 - B_2^2)B_c^{-2}. \quad (53)$$

Using Eq. (47), we obtain the indices of refraction for this case:

$$n_\pm = 1 + \{(\eta_1+\eta_2) \pm [(\eta_1-\eta_2)^2 +\eta_3^2]^{1/2}\} (B_1^2+B_2^2)B_c^{-2}. \tag{54}$$

The condition of no birefringence in Eq. (54) means that $[(\eta_1-\eta_2)^2 +\eta_3^2]$ vanishes, i.e.,

$$\eta_1 = \eta_2, \quad \eta_3 = 0, \text{ and no constraint on } \xi \tag{55}$$

This shows that Eq. (55) is a necessary condition for no birefringence. For $\mathbf{E}^{ext} = 0$, the refractive indices in the transverse external magnetic field $\mathbf{B}^{ext}$ for the linearly polarized lights whose polarizations are parallel and orthogonal to the magnetic field, are as follows:

$$n_\| = 1 + \{(\eta_1+\eta_2) + [(\eta_1-\eta_2)^2 +\eta_3^2]^{1/2}\} (\mathbf{B}^{ext})^2 B_c^{-2} \quad (\mathbf{E}^{wave} \| \mathbf{B}^{ext}), \tag{56}$$

$$n_\perp = 1 + \{(\eta_1+\eta_2) - [(\eta_1-\eta_2)^2 +\eta_3^2]^{1/2}\} (\mathbf{B}^{ext})^2 B_c^{-2} \quad (\mathbf{E}^{wave} \perp \mathbf{B}^{ext}). \tag{57}$$

For $\mathbf{B}^{ext} = 0$, we derive in the following the refractive indices in the transverse external electric field $\mathbf{E}^{ext}$ for the linearly polarized lights whose polarizations (electric fields) are parallel and orthogonal to the electric field. First, we use (39)-(41) & (43)-(46) to obtain

$$\varepsilon_{\alpha\beta} = \delta_{\alpha\beta}[1+2\eta_1 \mathbf{E}^2 B_c^{-2}] + 4\eta_1 E_\alpha E_\beta B_c^{-2}, \tag{58}$$

$$(\mu^{-1})_{\alpha\beta} = \delta_{\alpha\beta}[1+2\eta_1 \mathbf{E}^2 B_c^{-2}] - 4\eta_2 E_\alpha E_\beta B_c^{-2}, \tag{59}$$

$$\lambda_{\alpha\beta} = \delta_{\alpha\beta}[\Phi + \eta_3 \mathbf{E}^2 B_c^{-2}] + 2\eta_3 E_\alpha E_\beta B_c^{-2}, \tag{60}$$

$$J_1 = -4\eta_1 E_1^2 B_c^{-2} - 4\eta_2 E_2^2 B_c^{-2} - 4\eta_3 E_1 E_2 B_c^{-2}, \tag{61}$$

$$J_2 = -4\eta_1 E_2^2 B_c^{-2} - 4\eta_2 E_1^2 B_c^{-2} + 4\eta_3 E_1 E_2 B_c^{-2}, \tag{62}$$

$$J = -4\eta_1 E_1 E_2 B_c^{-2} + 4\eta_2 E_1 E_2 B_c^{-2} + 2\eta_3 (E_1^2 - E_2^2) B_c^{-2}. \tag{63}$$

Using (47), we obtain the indices of refraction for this case:

$$n_\pm = 1 + \{(\eta_1+\eta_2) \pm [(\eta_1-\eta_2)^2 +\eta_3^2]^{1/2}\} (E_1^2+E_2^2) B_c^{-2}. \tag{64}$$

The condition of no birefringence in (64) is the same as (55), i.e., that $[(\eta_1-\eta_2)^2 +\eta_3^2]$ vanishes. For $\mathbf{B}^{ext} = 0$, the refractive indices in the transverse external magnetic field $\mathbf{E}^{ext}$ for the linearly polarized lights whose polarizations are parallel and orthogonal to the magnetic field, are as follows:

$$n_\| = 1 + \{(\eta_1+\eta_2) + [(\eta_1-\eta_2)^2 +\eta_3^2]^{1/2}\} (\mathbf{E}^{ext})^2 B_c^{-2} \quad (\mathbf{E}^{wave} \| \mathbf{E}^{ext}), \tag{65}$$

$$n_\perp = 1 + \{(\eta_1+\eta_2) - [(\eta_1-\eta_2)^2 + \eta_3^2]^{1/2}\} (E^{ext})^2 B_c^{-2} \quad (E^{wave} \perp E^{ext}). \tag{66}$$

The magnetic field near pulsars can reach $10^{12}$ G, while the magnetic field near magnetars can reach $10^{15}$ G. The astrophysical processes in these locations need nonlinear electrodynamics to model. In the following section, we turn to experiments to measure the parameters of the PPM electrodynamics.

## 4. Measuring the parameters of the PPM electrodynamics

There are four parameters $\eta_1$, $\eta_2$, $\eta_3$, and $\xi$ in PPM electrodynamics to be measured by experiments. For the QED (Quantum Electrodynamics) corrections to classical electrodynamics, $\eta_1 = \alpha/(45\pi) = 5.1 \times 10^{-5}$, $\eta_2 = 7\alpha/(180\pi) = 9.0 \times 10^{-5}$, $\eta_3 = 0$, and $\xi = 0$. There are three vacuum birefringence experiments on going in the world to measure this QED vacuum birefringence – the BMV experiment (Battesti et al., 2008), the PVLAS experiment (Zavattini et al., 2008) and the Q & A experiment (Chen et al., 2007; Mei et al., 2010). The birefringence $\Delta n$ in the QED vacuum birefringence in a magnetic field $\mathbf{B}^{ext}$ is

$$\Delta n = n_\parallel - n_\perp = 4.0 \times 10^{-24} (\mathbf{B}^{ext}/1T)^2. \tag{67}$$

For 2.3 T field of the Q & A rotating permanent magnet, $\Delta n$ is $2.1 \times 10^{-23}$. This is about the same order of magnitude change in fractional length that ground interferometers for gravitational-wave detection aim at. Quite a lot of techniques developed in the gravitational-wave detection community are readily applicable for vacuum birefringence detection (Ni et al., 1991).

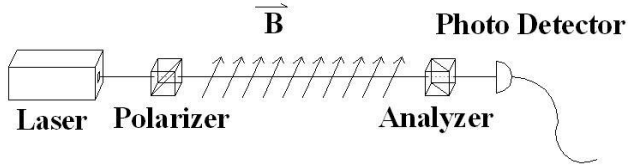

Fig. 2. Principle of vacuum dichroism and birefringence measurement.

The basic principle of these experimental measurements is shown as Fig. 2. The laser light goes through a polarizer and becomes polarized. This polarized light goes through a region of magnetic field. Its polarization status is subsequently analyzed by the analyzer-detector subsystem to extract the polarization effect imprinted in the region of the magnetic field. Since the polarization effect of vacuum birefringence in the magnetic field that can be produced on earth is extremely small, one has to multiply the optical pass through the magnetic field by using reflections or Fabry-Perot cavities. An already performed experiment, the BFRT experiment (Cameron et al., 1993) used multiple reflections; PVLAS, Q & A, BMV experiments all use Fabry-Perot cavities. For polarization experiment, Fabry-Perot cavity has the advantage of normal incidence of laser light which suppressed the part of polarization due to slant angle of reflections. With Fabry-Perot cavity, one needs to control the laser frequency and/or the cavity length so that the cavity is in resonance. With a finesse of 30,000, the resonant width (FWHM) is 17.7 pm for light with 1064 nm wavelength; when rms cavity length control is 10 % of this width, the precision would be 2.1 pm. Hence, one needs a feedback mechanism to lock the cavity to the laser or vice versa. For this, a commonly used scheme is Pound-Drever-Hall method (Drever et al., 1983). Vibration introduces noises in the Fabry-Perot cavity mirrors and hence, in the light intensity and light polarization transmitted through the Fabry-Perot cavity. Since the analyzer-detector subsystem detects light intensity to deduce the polarization effect, both intensity noise and polarization noise will contribute to the measurement results. Gravitational-wave community has a long-standing R & D on this. We benefit from their research advancements.

Now we illustrate with our Q & A experiment. Since 1991 we have worked on precision interferometry --- laser stabilization schemes, laser metrology and Fabry-Perot interferometers. With these experiences, we started in 1994 to build a 3.5 m prototype interferometer for measuring vacuum birefringence and improving the sensitivity of axion search as part of our continuing effort in precision interferometry. In 2002, we finished Phase I of constructing the 3.5 m prototype interferometer and made some Cotton-Mouton coefficient and Verdet coefficient measurements with a 1T electromagnet (Wu et al., 2002). The two vacuum tanks shown on the left photo of Fig. 3 house the two 5 cm-diameter Pabry-Perot mirrors with their suspensions; the 1T electromagnet had been in place of permanent magnet in the middle of the photo.

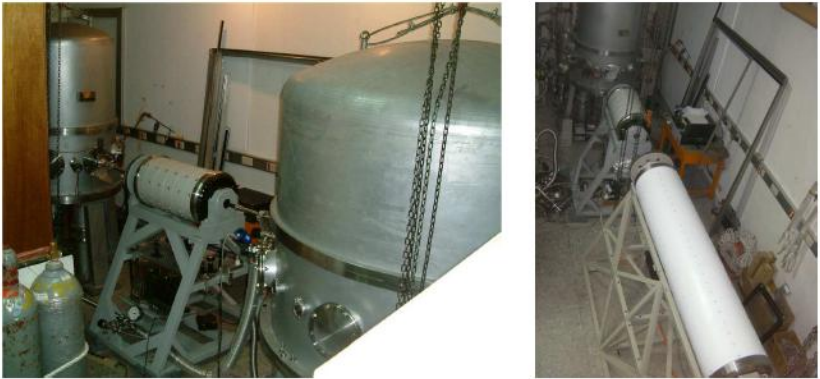

Fig. 3. Photo on the left-hand side shows the Q & A apparatus for Phase II experiment; photo on the right-hand side shows the Q & A apparatus for Phase III experiment.

Starting 2002, we had been in Phase II of Q & A experiment until 2008. The results of Phase II on dichroism and Cotton-Mouton effect measurement had been reported (Chen et al., 2007; Mei et al. 2009). At the end of Phase II, our sensitivity was still short from detection of QED vacuum birefringence by 3 orders of magnitude; so was the PVLAS experiment and had been the BFRT experiment. In 2009, we started Phase III of the Q & A experiment to extend the 3.5 m interferometer to 7 m with various upgrades. Photo on the left of Fig. 3 shows the apparatus for Phase II; photo on the right side of Fig. 3 shows the apparatus for Phase III, with the big (front) tank moved further to the front (out of the photo). The laser has been changed to 532 nm wavelength and is located next and beyond the front tank. We have installed a new 1.8 m 2.3 T permanent magnet (in the middle to bottom of right side photo) capable of rotation up to 13 cycles per second to enhance the physical effects. We are working with 532 nm Nd:YAG laser as light source with cavity finesse around 100,000, and aim at 10 nrad(Hz)$^{-1/2}$ optical sensitivity. With all these achieved and the upgrading of vacuum, for a period of 50 days (with duty cycle around 78 % as performed before) the vacuum birefringence measurement would be improved in precision by 3-4 orders of magnitude, and QED birefringence would be measured to 28 % (Mei et al., 2010). To enhance the physical effects further, another 1.8 m magnet will be added in the future.

All three ongoing experiments – PVLAS, Q &A, and BMV – are measuring the birefringence $\Delta n$, and hence, $\eta_1-\eta_2$ in case $\eta_3$ is assumed to be zero. To measure $\eta_1$ and $\eta_2$ separately, one-arm common path polarization measurement interferometer is not enough. We need a two-arm interferometer with the paths in two arms in magnetic fields with different strengths (or one with no magnetic field).

To measure $\eta_3$ in addition, one needs to use both external electric and external magnetic field. One possibility is to let light goes through strong microwave cavity and interferes. Suppose light propagation direction is the same as the microwave propagation direction which is perpendicular to the microwave fields. Let's choose $z$-axis to be in the propagation direction, $x$-axis in the $\mathbf{E}^{ext}$ direction and $y$-axis in the $\mathbf{B}^{ext}$ direction, i.e., $\mathbf{k} = (0, 0, k)$, $\mathbf{E}^{ext} = (E, 0, 0)$ and $\mathbf{B}^{ext} = (0, B, 0)$. We calculate the indices of refraction using Eqs. (39)-(47) without first assuming $E = B$ and obtain the following

$$\varepsilon_{\alpha\beta}: \varepsilon_{11}=1+2\eta_1(E^2-B^2)B_c^{-2}+4\eta_1 E^2 B_c^{-2}; \quad \varepsilon_{22}=1+2\eta_1(E^2-B^2)B_c^{-2}+4\eta_2 B^2 B_c^{-2};$$
$$\varepsilon_{33}=1+2\eta_1(E^2-B^2)B_c^{-2}; \quad \varepsilon_{12}=\varepsilon_{21}=2\eta_3 EB B_c^{-2}; \quad \varepsilon_{13}=\varepsilon_{23}=\varepsilon_{31}=\varepsilon_{32}=0, \quad (68)$$

$$(\mu^{-1})_{\alpha\beta}: (\mu^{-1})_{11}=1+2\eta_1(E^2-B^2)B_c^{-2}-4\eta_2 E^2 B_c^{-2}; \quad (\mu^{-1})_{22}=1+2\eta_1(E^2-B^2)B_c^{-2}-4\eta_1 B^2 B_c^{-2};$$
$$(\mu^{-1})_{33}=1+2\eta_1(E^2-B^2)B_c^{-2}; \quad (\mu^{-1})_{12}=(\mu^{-1})_{21}=2\eta_3 EB B_c^{-2}; \quad (\mu^{-1})_{13}=(\mu^{-1})_{23}=(\mu^{-1})_{13}=(\mu^{-1})_{23}=0, \quad (69)$$

$$\lambda_{\alpha\beta}: \lambda_{11}=\xi\Phi+\eta_3(E^2-B^2)B_c^{-2}+2\eta_3 E_2 B_c^{-2}; \quad \lambda_{22}=\xi\Phi+\eta_3(E^2-B^2)B_c^{-2}+2\eta_3 B^2 B_c^{-2};$$
$$\lambda_{33}=\xi\Phi+\eta_3(E^2-B^2)B_c^{-2}; \quad \lambda_{12}=-4\eta_1 EB B_c^{-2}; \quad \lambda_{21}=4\eta_2 EB B_c^{-2}; \quad \lambda_{13}=\lambda_{23}=\lambda_{31}=\lambda_{32}=0, \quad (70)$$

$$J_1 \equiv -4\eta_1(E^2+B^2)B_c^{-2}+4\eta_1 EB B_c^{-2}, \quad (71)$$

$$J_2 \equiv -4\eta_2(E^2+B^2)B_c^{-2}+4\eta_2 EB B_c^{-2}, \quad (72)$$

$$J \equiv 2\eta_3(E^2-B^2)B_c^{-2}, \quad (73)$$

$$n_\pm = 1 + (\eta_1+\eta_2)(E^2+B^2-EB)B_c^{-2} \pm [(\eta_1-\eta_2)^2(E^2+B^2-EB)^2+\eta_3^2(E^2-B^2)]^{1/2} B_c^{-2}. \quad (74)$$

As a consistent check, there is no birefringence in Eq. (74) for $\eta_1 = \eta_2$, $\eta_3 = 0$.

Now, we consider two special cases for Eq. (74): (i) $E=B$ as in the strong microwave cavity, the indices of refraction for light is

$$n_\pm = 1 + (\eta_1+\eta_2)B^2 B_c^{-2} \pm (\eta_1-\eta_2)B^2 B_c^{-2}, \quad (75)$$

with birefringence Δn given by

$$\Delta n = 2(\eta_1-\eta_2)B^2B_c^{-2}; \tag{76}$$

(ii) $E=0$, $B\neq 0$, the indices of refraction for light is

$$n_\pm = 1 + (\eta_1+\eta_2)B^2B_c^{-2} \pm [(\eta_1-\eta_2)^2+\eta_3^2]^{1/2}B^2B_c^{-2}, \tag{77}$$

with birefringence Δn given by

$$\Delta n = 2[(\eta_1-\eta_2)^2+\eta_3^2]^{1/2}B^2B_c^{-2}. \tag{78}$$

Equation (77) agrees with (54) derived earlier.

To measure $\eta_1$, $\eta_2$ and $\eta_3$, we could do the following three experiments to determine them: (i) to measure the birefringence $\Delta n = 2(\eta_1-\eta_2)B^2B_c^{-2}$ of light with the external field provided by a strong microwave cavity or wave guide to determine $\eta_1-\eta_2$; (ii) to measure the birefringence $\Delta n = 2[(\eta_1-\eta_2)^2+\eta_3^2]^{1/2}B^2B_c^{-2}$ of light with the external magnetic field provided by a strong magnet to determine $\eta_3$ with $\eta_1-\eta_2$ determined by (i); (iii) to measure $\eta_1$ and $\eta_2$ separately using two-arm interferometer with the paths in two arms in magnetic fields with different strengths (or one with no magnetic field).

As to the term $\xi\Phi$ and parameter $\xi$, it does not give any change in the index of refraction. However, as we will see in section 7 and section 8, it gives a polarization rotation and the effect can be measured though observations with astrophysical and cosmological propagation of electromagnetic waves.

## 5. Electromagnetism in curved spacetime and the Einstein equivalence principle

In the earth laboratory, where variation of gravity is small, we can use standard Maxwell equations together with Lorentz force law for ordinary measurements and experiments. However, in precision experiments on earth, in space, in the astrophysical situation or in the cosmological setting, the gravity plays an important role and is non-negligible. In the remaining part of this chapter, we address to the issue of electromagnetism in gravity and more empirical tests of electromagnetism and special relativity. The standard way of including gravitational effects in electromagnetism is to use the comma-goes-to-semicolon rule, i. e., the principle of equivalence (the minimal coupling rule). This is the essence of Einstein Equivalence Principle (EEP) which states that everywhere in the 4-dimensional spacetime, locally, the physics is that of special relativity. This guarantees that the 4-dimensional geometry can be described by a metric $g_{ij}$ which can be transformed into the Minkowski metric locally. In curved spacetime, $\eta_{ij}$ is replaced by $g_{ij}$ with partial derivative (comma) replaced by the covariant derivative in the $g_{ij}$ metric (semi-colon) in the Lagrangian density for a system of charged particles. When this is done the Lagrangian density becomes

$$L_I = -(1/(16\pi))\chi_{GR}{}^{ijkl} F_{ij} F_{kl} - A_k j^k (-g)^{(1/2)} - \Sigma_I m_I (ds_I)/(dt) \delta(\mathbf{x}-\mathbf{x}_I), \tag{79}$$

where the GR (General Relativity) constitutive tensor $\chi_{GR}^{ijkl}$ is given by

$$\chi_{GR}^{ijkl} = (-g)^{1/2} [(1/2)\, g^{ik} g^{jl} - (1/2)\, g^{il} g^{kj}]. \tag{80}$$

In general relativity or metric theories of gravity where EEP holds, the line element near a world point (event) $P$ is given by

$$ds^2 = g_{ij}\, dx^i\, dx^j = g_{AB}\, dx^A\, dx^B = [\eta_{AB} + O((\Delta x^C)^2)]\, dx^A\, dx^B, \tag{81}$$

where $\{x^i\}$ is an arbitrary coordinate system, $\{x^A\}$ is a locally inertial frame, and $g_{ij}$ & $g_{AB}$ are the metric tensor in their respective frames. According to the definition of locally inertial frame, we have $g_{AB} = \eta_{AB} + O((\Delta x^C)^2)$. Therefore, in the locally inertial system near $P$, special relativity holds up to the curvature ambiguity, and the definition of rods and clocks is the same as in the special relativity including local quantum mechanics and electromagnetism.

Nevertheless, for long range propagation and large-scale phenomenon, curvature effects are important. For long range electromagnetic propagation, wavelength/frequency shift is important. From distant quasars, the redshift factor z exceeds 6, i.e., the wavelength changes by more than 6-fold. The gravitational redshift is given by

$$\Delta\tau_A/\Delta\tau_B = g_{00}(B)/g_{00}(A), \tag{82}$$

where $\Delta\tau_A$ and $\Delta\tau_B$ are the proper periods of a light signal emitted by a source A and received by B respectively. This formula applies equally well to the solar system, to galaxies and to cosmos. Its realm of practical application is in clock and frequency comparisons. In the weak gravitational field such as near earth or in the solar system, we have

$$g_{00} = 1 - 2U/c^2, \tag{83}$$

in the first approximation, where $U$ is the Newtonian potential. On the surface of earth, $U/c^2 \approx 0.7 \times 10^{-9}$ and the redshift is a fraction of it. This redshift is measured in the laboratory and in space borne missions. It is regularly corrected for the satellite navigation systems such as GPS, GLONASS, Galileo and Beidou. Another effect of electromagnetic propagation in gravity is its deflection with important application to gravitational lensing effects in astrophysics.

## 6. Empirical tests of electromagnetism in gravity and the χ-g framework

In section 1, we have discussed the constraints on Proca part of Lagrangian density, i.e., photon mass. In this section, we discuss the empirical foundation of the Maxwell (main) part of electromagnetism. First we need a framework to interpret experimental tests. A natural framework is to extend the GR constitutive tensor $\chi_{GR}^{ijkl}$ [equation (80)] into a general form, and look for experimental and observational evidences to test it to see how much it is constrained to the GR form. The general framework we adopt is the *χ-g* framework (Ni, 1983a, 1984a, 1984b, 2010).

The *χ-g* framework can be summarized in the following interaction Lagrangian density

$$L_I = -(1/(16\pi))\chi^{ijkl} F_{ij} F_{kl} - A_k j^k (-g)^{(1/2)} - \Sigma_I m_I (ds_I)/(dt) \delta(\boldsymbol{x}-\boldsymbol{x}_I), \tag{84}$$

where $\chi^{ijkl} = \chi^{klij} = -\chi^{jikl}$ is a tensor density of the gravitational fields (e.g., $g_{ij}$, $\varphi$, etc.) or fields to be investigated and $F_{ij} \equiv A_{j,i} - A_{i,j}$ etc. have the usual meaning in classical electromagnetism. The gravitation constitutive tensor density $\chi^{ijkl}$ dictates the behaviour of electromagnetism in a gravitational field and has 21 independent components in general. For general relativity or a metric theory (when EEP holds), $\chi^{ijkl}$ is determined completely by the metric $g_{ij}$ and equals $(-g)^{1/2} [(1/2) g^{ik} g^{jl} - (1/2) g^{il} g^{jk}]$.

In the following, we use experiments and observations to constrain the 21 degrees of freedom of $\chi^{ijkl}$ to see how close we can reach general relativity. This procedure also serves to reinforce the empirical foundations of classical electromagnetism as EEP locally is based on special relativity including classical electromagnetism.

In the $\chi$-g framework, for a weak gravitational field,

$$\chi^{ijkl} = \chi^{(0)ijkl} + \chi^{(1)ijkl}, \tag{85}$$

where

$$\chi^{(0)ijkl} = (1/2)\eta^{ik}\eta^{jl} - (1/2)\eta^{il}\eta^{kj}, \tag{86}$$

with $\eta^{ij}$ the Minkowski metric and $|\chi^{(1)ijkl}| \ll 1$ for all $i$, $j$, $k$, and $l$. The small special relativity violation (constant part), if any, is put into the $\chi^{(1)ijkl}$'s. In this field the dispersion relation for $\omega$ for a plane-wave propagating in the z-direction is

$$\omega_{\pm} = k\{1+(1/4)[(K_1+K_2) \pm [(K_1-K_2)^2 + 4 K^2]^{1/2}]\}, \tag{87}$$

where

$$K_1 = \chi^{(1)1010} - 2\chi^{(1)1013} + \chi^{(1)1313}, \tag{88}$$

$$K_2 = \chi^{(1)2020} - 2\chi^{(1)2023} + \chi^{(1)2323}, \tag{89}$$

$$K = \chi^{(1)1020} - \chi^{(1)1023} - \chi^{(1)1320} + \chi^{(1)1323}. \tag{90}$$

Photons with two different polarizations propagate with different speeds $V_{\pm} = \omega_{\pm}/k$ and would split in 4-dimensional spacetime. The conditions for no splitting (no retardation) is $\omega_+ = \omega_-$, i.e.,

$$K_1 = K_2, \qquad K = 0. \tag{91}$$

Eq. (91) gives two constraints on the $\chi^{(1)ijkl}$'s (Ni, 1983a, 1984a, 1984b).

*Constraints from no birefringence.* The condition for no birefringence (no splitting, no retardation) for electromagnetic wave propagation in all directions in the weak field limit gives ten independent constraint equations on the constitutive tensor $\chi^{ijkl}$'s. With these ten constraints, the constitutive tensor $\chi^{ijkl}$ can be written in the following form

$$\chi^{ijkl} = (-H)^{1/2}[(1/2)H^{ik}H^{jl} - (1/2)H^{il}H^{kj}]\psi + \varphi e^{ijkl}, \tag{92}$$

where $H = \det(H_{ij})$ and $H_{ij}$ is a metric which generates the light cone for electromagnetic propagation (Ni, 1983a, 1984a,b). Note that (92) has an axion degree of freedom, $\varphi e^{ijkl}$, and a 'dilaton' degree of freedom, $\psi$. Lämmerzahl and Hehl (2004) have shown that this non-birefringence guarantees, without approximation, Riemannian light cone, i.e., Eq. (92) holds without the assumption of weak field also. To fully recover EEP, we need (i) good constraints from no birefringence, (ii) good constraints on no extra physical metric, (iii) good constraints on no $\psi$ ('dilaton'), and (iv) good constraints on no $\varphi$ (axion) or no pseudoscalar-photon interaction.

Eq. (92) is verified empirically to high accuracy from pulsar observations and from polarization measurements of extragalactic radio sources. With the null-birefringence observations of pulsar pulses and micropulses before 1980, the relations (92) for testing EEP are empirically verified to $10^{-14} – 10^{-16}$ (Ni, 1983a, 1984a, 1984b). With the present pulsar observations, these limits would be improved; a detailed such analysis is given by Huang (2002). Analyzing the data from polarization measurements of extragalactic radio sources, Haugan and Kauffmann (1995) inferred that the resolution for null-birefringence is 0.02 cycle at 5 GHz. This corresponds to a time resolution of $4 \times 10^{-12}$ s and gives much better constraints. With a detailed analysis and more extragalactic radio observations, (92) would be tested down to $10^{-28}$-$10^{-29}$ at cosmological distances. In 2002, Kostelecky and Mews (2002) used polarization measurements of light from cosmologically distant astrophysical sources to yield stringent constraints down to $2 \times 10^{-32}$. For a review, see Ni (2010). In the remaining part of this subsection, we assume (92) to be correct.

*Constraints on one physical metric and no 'dilaton' ($\psi$).* Let us now look into the empirical constraints for $H^{ij}$ and $\psi$. In Eq. (85), $ds$ is the line element determined from the metric $g_{ij}$. From Eq. (92), the gravitational coupling to electromagnetism is determined by the metric $H_{ij}$ and two (pseudo)scalar fields $\varphi$ 'axion' and $\psi$ 'dilaton'. If $H_{ij}$ is not proportional to $g_{ij}$, then the hyperfine levels of the lithium atom, the beryllium atom, the mercury atom and other atoms will have additional shifts. But this is not observed to high accuracy in Hughes-Drever experiments (Hughes et al., 1960; Beltran-Lopez et al., 1961; Drever, 1961; Ellena et al., 1987; Chupp et al., 1989). Therefore $H_{ij}$ is proportional to $g_{ij}$ to certain accuracy. Since a change of $H^{ik}$ to $\lambda H^{ij}$ does not affect $\chi^{ijkl}$ in Eq. (92), we can define $H_{11} = g_{11}$ to remove this scale freedom (Ni, 1983a, 1984a). For a review, see Ni (2010).

Eötvös-Dicke experiments (Eötvös, 1890; Eötvös et al., 1922; Roll et al., 1964; Braginsky and Panov, 1971; Schlamminger et al., 2008 and references therein) are performed on unpolarized test bodies. In essence, these experiments show that unpolarized electric and magnetic energies follow the same trajectories as other forms of energy to certain accuracy. The constraints on Eq. (92) are

$$| 1-\psi | / U < 10^{-10}, \tag{93}$$

and

$$| H_{00} - g_{00} | / U < 10^{-6}, \qquad (94)$$

where $U$ (~ $10^{-6}$) is the solar gravitational potential at the earth.

In 1976, Vessot et al. (1980) used an atomic hydrogen maser clock in a space probe to test and confirm the metric gravitational redshift to an accuracy of $1.4 \times 10^{-4}$, i. e.,

$$| H_{00} - g_{00} | / U \leq 1.4 \times 10^{-4}, \qquad (95)$$

where $U$ is the change of earth gravitational field that the maser clock experienced.

With constraints from (i) no birefringence, (ii) no extra physical metric, (iii) no $\psi$ ('dilaton'), we arrive at the theory (84) with $\chi^{ijkl}$ given by

$$\chi^{ijkl} = (-g)^{1/2} [(1/2) g^{ik} g^{jl} - (1/2) g^{il} g^{kj} + \varphi \, \varepsilon^{ijkl}], \qquad (96)$$

i.e., an axion theory (Ni, 1983a, 1984a; Hehl and Obukhov 2008). Here $\varepsilon^{ijkl}$ is defined to be $(-g)^{-1/2} e^{ijkl}$. The current constraints on $\varphi$ from astrophysical observations and CMB polarization observations will be discussed in section 8. Thus, from experiments and observations, only one degree of freedom of $\chi^{ijkl}$ is not much constrained.

Now let's turn into more formal aspects of equivalence principles. We proved that for a system whose Lagrangian density is given by Eq. (84), the Galileo Equivalence Principle (UFF [Universality of Free Fall; WEP I) holds if and only if Eq. (96) holds (Ni, 1974, 1977).

If $\varphi \neq 0$ in (96), the gravitational coupling to electromagnetism is not minimal and EEP is violated. Hence WEP I does not imply EEP and Schiff's conjecture (which states that WEP I implies EEP) is incorrect (Ni, 1973, 1974, 1977). However, WEP I does constrain the 21 degrees of freedom of χ to only one degree of freedom (φ), and Schiff's conjecture is largely right in spirit.

The theory with $\varphi \neq 0$ is a pseudoscalar theory with important astrophysical and cosmological consequences (section 8). This is an example that investigations in fundamental physical laws lead to implications in cosmology. Investigations of CP problems in high energy physics leads to a theory with a similar piece of Lagrangian with φ the axion field for QCD (Peccei and Quinn, 1977; Weinberg, 1978; Wilczek, 1978).

In the nonmetric theory with $\chi^{ijkl}$ ($\varphi \neq 0$) given by Eq. (96) (Ni 1973, 1974, 1977), there are anomalous torques on electromagnetic-energy-polarized bodies so that different test bodies will change their rotation state differently, like magnets in magnetic fields. Since the motion of a macroscopic test body is determined not only by its trajectory but also by its rotation state, the motion of polarized test bodies will not be the same. We, therefore, have proposed the following stronger weak equivalence principle (WEP II) to be tested by experiments, which states that in a gravitational field, both the translational and rotational motion of a test body with a given initial motion state is independent of its internal structure and composition (universality of free-fall motion) (Ni 1974, Ni 1977). To put in another way, the behavior of motion including rotation is that in a local inertial frame for test-bodies. If WEP II is violated, then EEP is violated. Therefore from above, in the χ-g framework, the imposition of WEP II guarantees that EEP is valid.

WEP II state that the motion of all six degrees of freedom (3 translational and 3 rotational) must be the same for all test bodies as in a local inertial frame. There are two different scenarios that WEP II would be violated: (i) the translational motion is affected by the rotational state; (ii) the rotational state changes with angular momentum (rotational direction/speed) or species. Recent experimental results of Gravity Probe B experiment with rotating quartz balls in earth orbit (Everitt et al., 2011) not just verifies frame-dragging effect, but also verifies both aspects of WEP II for unpolarized-bodies to an ultimate precision (Ni, 2011).

In this section, we have shown that the empirical foundation of classical electromagnetism is solid except in the aspect of a pseudoscalar photon interaction. This exception has important consequences in cosmology. In the following two sections, we address this issue.

## 7. Pseudoscalar-photon interaction

In this section, we discuss the modified electromagnetism in gravity with the pseudoscalar-photon interaction which was reached in the last section, i.e., the theory with the constitutive tensor density (96). Its Lagrangian density is as follows

$$L_I = - (1/(16\pi))(-g)^{1/2}[(1/2)g^{ik}g^{jl}-(1/2)g^{il}g^{kj}+\varphi\, \varepsilon^{ijkl}]F_{ij}F_{kl} - A_k j^k (-g)^{(1/2)} - \Sigma_I m_I(ds_I)/(dt)\delta(\boldsymbol{x}-\boldsymbol{x}_I). \quad (97)$$

In the constitutive tensor density and the Lagrangian density, $\varphi$ is a scalar or pseudoscalar function of relevant variables. If we assume that the $\varphi$-term is local CPT invariant, then $\varphi$ should be a pseudoscalar (function) since $\varepsilon^{ijkl}$ is a pseudotensor. The pseudoscalar(scalar)-photon interaction part (or the nonmetric part) of the Lagrangian density of this theory is

$$L^{(\varphi\gamma\gamma)} = L^{(NM)} = - (1/16\pi)\, \varphi\, e^{ijkl}F_{ij}F_{kl} = - (1/4\pi)\, \varphi_{,i}\, e^{ijkl}A_j A_{k,l} \text{ (mod div)}, \quad (98)$$

where 'mod div' means that the two Lagrangian densities are related by integration by parts in the action integral. This term gives pseudoscalar-photon-photon interaction in the quantum regime and can be denoted by $L^{(\varphi\gamma\gamma)}$. This term is also the $\xi$-term in the PPM Lagrangian density $L_{PPM}$ with the $\varphi \equiv (1/4)\xi\Phi$ correspondence. The Maxwell equations (Ni 1973, 1977) from Eq. (98) become

$$F^{ik}{}_{;k} + \varepsilon^{ikml} F_{km}\varphi_{,l} = -4\pi j^i, \quad (99)$$

where the derivation $_;$ is with respect to the Christoffel connection of the metric. The Lorentz force law is the same as in metric theories of gravity or general relativity. Gauge invariance and charge conservation are guaranteed. Related to charge consevation tests, please see Lämmerzahl et al. (2007). The modified Maxwell equations are also conformally invariant.

The rightest term in equation (99) is reminiscent of Chern-Simons (1974) term $e^{\alpha\beta\gamma} A_\alpha F_{\beta\gamma}$. There are two differences: (i) Chern-Simons term is in 3 dimensional space; (ii) Chern-Simons term in the integral is a total divergence (Table 1). However, it is interesting to notice that the cosmological time may be defined through the Chern-Simons invariant (Smolin and Soo, 1995).

| Term | Dimension | Reference | Meaning |
|---|---|---|---|
| $e^{\alpha\beta\gamma} A_\alpha F_{\beta\gamma}$ | 3 | Chern-Simons (1974) | Intergrand for topological invarinat |
| $e^{ijkl} \varphi F_{ij} F_{kl}$ | 4 | Ni (1973, 1974, 1977) | Pseudoscalar-photon coupling |
| $e^{ijkl} \varphi F^{QCD}{}_{ij} F^{QCD}{}_{kl}$ | 4 | Peccei-Quinn (1977) Weinberg (1978) Wilczek (1978) | Pseudoscalar-gluon coupling |
| $e^{ijkl} V_i A_j F_{kl}$ | 4 | Carroll-Field-Jackiw (1990) | External constant vector coupling |

**Table 1.** Various terms in the Lagrangian and their meaning.

A term similar to the one in equation (99) (axion-gluon interaction) occurs in QCD in an effort to solve the strong CP problem (Peccei & Quinn, 1977; Weinberg, 1978; Wilczek, 1978). Carroll, Field and Jackiw (1990) proposed a modification of electrodynamics with an additional $e^{ijkl} V_i A_j F_{kl}$ term with $V_i$ a constant vector (See also Jackiw, 2007). This term is a special case of the term $e^{ijkl} \varphi F_{ij} F_{kl}$ (mod div) with $\varphi_{,i} = -\tfrac{1}{2} V_i$.

Various terms in the Lagrangians discussed in this subsection are listed in Table 1. Empirical tests of the pseudoscalar-photon interaction (99) will be discussed in next section.

## 8. Cosmic polarization rotation

For the electromagnetism in gravity with an effective pseudoscalar-photon interaction discussed in the last section, the electromagnetic wave propagation equation is governed by equation (99). In a local inertial (Lorentz) frame of the $g$-metric, it is reduced to

$$F^{ik}{}_{,k} + e^{ikml} F_{km} \varphi_{,l} = 0. \tag{100}$$

Analyzing the wave into Fourier components, imposing the radiation gauge condition, and solving the dispersion eigenvalue problem, we obtain $k = \omega + (n^\mu \varphi_{,\mu} + \varphi_{,0})$ for right circularly polarized wave and $k = \omega - (n^\mu \varphi_{,\mu} + \varphi_{,0})$ for left circularly polarized wave in the eikonal approximation (Ni 1973). Here $n^\mu$ is the unit 3-vector in the propagation direction. The group velocity is

$$v_g = \partial\omega/\partial k = 1, \tag{101}$$

which is independent of polarization. There is no birefringence. For the right circularly polarized electromagnetic wave, the propagation from a point $P_1 = \{x_{(1)}{}^i\} = \{x_{(1)}{}^0; x_{(1)}{}^\mu\} = \{x_{(1)}{}^0, x_{(1)}{}^1, x_{(1)}{}^2, x_{(1)}{}^3\}$ to another point $P_2 = \{x_{(2)}{}^i\} = \{x_{(2)}{}^0; x_{(2)}{}^\mu\} = \{x_{(2)}{}^0, x_{(2)}{}^1, x_{(2)}{}^2, x_{(2)}{}^3\}$ adds a phase of $a = \varphi(P_2) - \varphi(P_1)$ to the wave; for left circularly polarized light, the added phase will be opposite in sign (Ni 1973). Linearly polarized electromagnetic wave is a superposition of circularly polarized waves. Its polarization vector will then rotate by an angle $a$. Locally, the polarization rotation angle can be approximated by

$$a = \varphi(P_2) - \varphi(P_1) = \Sigma_{i=0}{}^3 [\varphi_{,i} \times (x_{(2)}{}^i - x_{(1)}{}^i)] = \Sigma_{i=0}{}^3 [\varphi_{,i} \Delta x^i] = \varphi_{,0} \Delta x^0 + [\Sigma_{\mu=1}{}^3 \varphi_{,\mu} \Delta x^\mu]$$
$$= \Sigma_{i=0}{}^3 [V_i \Delta x^i] = V_0 \Delta x^0 + [\Sigma_{\mu=1}{}^3 V_\mu \Delta x^\mu]. \tag{102}$$

The rotation angle in (102) consists of 2 parts -- $\varphi_{,0}\Delta x^0$ and $[\Sigma_{\mu=1}^3 \varphi_{,\mu}\Delta x^\mu]$. For light in a local inertial frame, $|\Delta x^\mu| = |\Delta x^0|$. In Fig. 4, space part of the rotation angle is shown. The amplitude of the space part depends on the direction of the propagation with the tip of magnitude on upper/lower sphere of diameter $|\Delta x^\mu| \times |\varphi_{,\mu}|$. The time part is equal to $\Delta x^0 \varphi_{,0}$. ($\nabla\varphi \equiv [\varphi_{,\mu}]$) When we integrate along light (wave) trajectory in a global situation, the total polarization rotation (relative to no $\varphi$-interaction) is again $\Delta\varphi = \varphi_2 - \varphi_1$ for $\varphi$ is a scalar field where $\varphi_1$ and $\varphi_2$ are the values of the scalar field at the beginning and end of the wave. When the propagation distance is over a large part of our observed universe, we call this phenomenon cosmic polarization rotation (Ni, 2008, 2010).

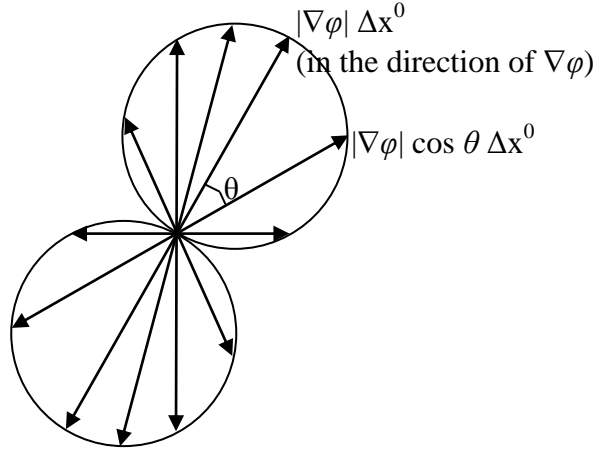

Fig. 4. Space contribution to the local polarization rotation angle -- $[\Sigma_{\mu=1}^3 \varphi_{,\mu}\Delta x^\mu] = |\nabla\varphi| \cos\theta \Delta x^0$. The time contribution is $\varphi_{,0} \Delta x^0$. The total contribution is $(|\nabla\varphi| \cos\theta + \varphi_{,0}) \Delta x^0$. ($\Delta x^0 > 0$).

In the CMB polarization observations, there are variations and fluctuations. The variations and fluctuations due to scalar-modified propagation can be expressed as $\delta\varphi(2) - \delta\varphi(1)$, where 2 denotes a point at the last scattering surface in the decoupling epoch and 1 observation point. $\delta\varphi(2)$ is the variation/fluctuation at the last scattering surface. $\delta\varphi(1)$ at the present observation point is zero or fixed. Therefore the covariance of fluctuation $<[\delta\varphi(2) - \delta\varphi(1)]^2>$ gives the covariance of $\delta\varphi^2(2)$ at the last scattering surface. Since our Universe is isotropic to ~ $10^{-5}$, this covariance is ~ $(\varsigma \times 10^{-5})^2$ where the parameter $\varsigma$ depends on various cosmological models. (Ni 2008, 2010)

Now we must say something about nomenclature.

Birefringence, also called double refraction, refers to the two different directions of propagation that a given incident ray can take in a medium, depending on the direction of polarization. The index of refraction depends on the direction of polarization.

Dichroic materials have the property that their absorption constant varies with polarization. When polarized light goes through dichroic material, its polarization is rotated due to difference in absorption in two principal directions of the material for the two polarization components. This phenomenon or property of the medium is called dichroism.

In a medium with optical activity, the direction of a linearly polarized beam will rotate as it propagates through the medium. A medium subjected to magnetic field becomes optically active and the associated polarization rotation is called Faraday rotation.

Cosmic polarization rotation is neither dichroism nor birefringence. It is more like optical activity, with the rotation angle independent of wavelength. Conforming to the common usage in optics, one should not call it cosmic birefringence -- *a misnomer*.

Now we review and compile the constraints of various analyses from CMB polarization observations.

In 2002, DASI microwave interferometer observed the polarization of the cosmic background (Kovac et al., 2002). E-mode polarization is detected with 4.9 σ. The TE correlation of the temperature and E-mode polarization is detected at 95% confidence. This correlation is expected from the Raleigh scattering of radiation. However, with the (pseudo)scalar-photon interaction under discussion, the polarization anisotropy is shifted differently in different directions relative to the temperature anisotropy due to propagation; the correlation will then be downgraded. In 2003, from the first-year data (WMAP1), WMAP found that the polarization and temperature are correlated to more than 10 σ (Bennett *et al* 2003). This gives a constraint of about $10^{-1}$ for $\Delta\varphi$ (Ni, 2005a, 2005b).

Further results and analyses of CMB polarization observations came out after 2006. In Table 2, we update our previous compilations (Ni 2008, 2010). Although these results look different at 1 σ level, they are all consistent with null detection and with one another at 2 σ level.

| Analysis | Constraint [mrad] | Source data |
|---|---|---|
| Ni (2005a, b) | ±100 | WMAP1 (Bennett *et al* 2003) |
| Feng, Li, Xia, Chen & Zhang (2006) | -105 ± 70 | WMAP3 (Spergel *et al* 2007) & BOOMERANG (B03) (Montroy *et al* 2006) |
| Liu, Lee & Ng (2006) | ±24 | BOOMERANG (B03) (Montroy *et al* 2006) |
| Kostelecky & Mews (2007) | 209 ± 122 | BOOMERANG (B03) (Montroy *et al* 2006) |
| Cabella, Natoli & Silk (2007) | -43 ± 52 | WMAP3 (Spergel *et al* 2007) |
| Xia, Li, Wang & Zhang (2008) | -108 ± 67 | WMAP3 (Spergel *et al* 2007) & BOOMERANG (B03) (Montroy *et al* 2006) |
| Komatsu *et al* (2009) | -30 ± 37 | WMAP5 (Komatsu *et al* 2009) |
| Xia, Li, Zhao & Zhang (2008) | -45 ± 33 | WMAP5 (Komatsu *et al* 2009) & BOOMERANG (B03) (Montroy *et al* 2006) |
| Kostelecky & Mews (2008) | 40 ± 94 | WMAP5 (Komatsu *et al* 2009) |
| Kahniashvili, Durrer & Maravin (2008) | ± 44 | WMAP5 (Komatsu *et al* 2009) |
| Wu *et al* (2009) | 9.6 ± 14.3 ± 8.7 | QuaD (Pryke *et al* 2009) |
| Brown *et al.* (2009) | 11.2 ± 8.7 ± 8.7 | QuaD (Brown *et al* 2009) |
| Komatsu *et al.* (2011) | -19 ± 22 ± 26 | WMAP7 (Komatsu *et al* 2011) |

Table 2. Constraints on cosmic polarization rotation from CMB polarization observations. [See Ni (2010) for detailed references.]

Both magnetic field and potential new physics affect the propagation of CMB propagation and generate BB power spectra from EE spectra of CMB. The Faraday rotation due to magnetic field is wavelength dependent while the cosmic polarization rotation due to effective pseudoscalar-photon interaction is wavelength-independent. This property can be used to separate the two effects. With the tensor mode generated by these two effects measured and subtracted, the remaining tensor mode perturbations could be analyzed for signals due to primordial (inflationary) gravitational waves (GWs). In Ni (2009a,b), we have discussed the direct detectability of these primordial GWs using space GW detectors.

Observations of radio and optical/UV polarization of radio galaxies are also sensitive to measure/test the cosmic polarization rotation, and give comparable constraints of several hundredths of an arcsecond. These observations have the capability of determining the polarization rotation in various directions. For a recent work, see di Serego Alighieri et al. (2010).

## 9. Outlook

We have looked at the foundations of electromagnetism in this chapter. For doing this, we have used two approaches. The first one is to formulate a Parametrized Post-Maxwellian framework to include QED corrections and a pseudoscalar photon interaction. We discuss various vacuum birefringence experiments – ongoing and proposed -- to measure these parameters. The second approach is to look at electromagnetism in gravity and various experiments and observations to determine its empirical foundation. We found that the foundation is solid with the only exception of a potentially possible pseudoscalar-photon interaction. We discussed its experimental constraints and look forward to more future experiments.

## 10. Acknowledgment

I would like to thank the National Science Council (Grants No. NSC97-2112-M-007-002 and No. NSC98-2112-M-007-009) for supporting this work in part.